\documentclass[runningheads]{llncs}
\usepackage{graphicx}
\usepackage{multirow}
\usepackage{amsmath, amssymb}
\usepackage{booktabs}
\usepackage{comment}
\usepackage{todonotes}
\usepackage{tabularx}

\begin{document}

\title{Self-Supervised Learning for Visual Summary Identification in Scientific Publications}

\titlerunning{SSL for Visual Summary Identification in Scientific Publications}
%
\author{Shintaro Yamamoto\inst{1} \and
Anne Lauscher\inst{2} \and
Simone Paolo Ponzetto\inst{2} \and \\
Goran Glava\v{s}\inst{2} \and
Shigeo Morishima\inst{3}}
\authorrunning{S. Yamamoto et al.}
%
\institute{Department of Pure and Applied Physics, Waseda University, Japan  \and
Data and Web Science Group, University of Mannheim, Germany \and
Waseda Research Institute for Science and Engineering, Japan \\
\email{s.yamamoto@fuji.waseda.jp}}
%
\maketitle              
\begin{abstract}
Providing visual summaries of scientific publications can increase information access for readers and thereby help deal with the exponential growth in the number of scientific publications. Nonetheless, efforts in providing visual publication summaries have been few and far apart, primarily focusing on the biomedical domain. This is primarily because of the limited availability of annotated gold standards, which hampers the application of robust and high-performing supervised learning techniques. To address these problems we create a new benchmark dataset for selecting figures to serve as visual summaries of publications based on their abstracts, covering several domains in computer science. Moreover, we develop a self-supervised learning approach, based on heuristic matching of inline references to figures with figure captions. Experiments in both biomedical and computer science domains show that our model is able to outperform the state of the art despite being self-supervised and therefore not relying on any annotated training data.

\end{abstract}
\setcounter{footnote}{0} 

\section{Introduction}
Given the exponential growth in the number of scientific publications~\cite{Bornmann2015}, providing concise summaries of scientific literature becomes increasingly important. Accordingly, previous work has focused on the automatic creation of textual summaries \cite{cohan-etal-2018-discourse,cohan-goharian-2015-scientific,Qiaozhu2008,Qazvinian2008,lauscher2017university,yasunaga2019scisummnet}.
However, specifically in the case of scientific publications (and especially in some domains), information is also conveyed in the form of figures, which allow the reader to understand the scientific contributions better, offering visual representations of data, experimental design, and results. Some scientific publishing companies (e.g., Elsevier) even require authors to submit a figure as a Graphical Abstract (GA), which is ``a single, concise, pictorial and visual summary of the main findings of the article''\footnote{\url{https://www.elsevier.com/authors/journal-authors/graphical-abstract}}. GAs, in turn, are then used to provide multi-modal online search results, following the observations that humans better remember and recall visual information \cite{nelson1976pictorial}.

Recently, Yang et al.\ \cite{Yang2019Identifying} introduced the concept of the central figure, referring to the figure that is the best candidate for a GA of a paper.
To build a dataset for automatically finding central figures in scientific publications, they asked authors of papers in PubMed\footnote{\url{https://pubmed.ncbi.nlm.nih.gov/}} to identify one central figure in each of their scientific publications.
Though a GA is not required for all publications, authors were shown to be able to identify the central figure from their publications for $87.6\%$ of the papers. Using the obtained datasets of publications with GAs, they devised a supervised machine learning approach for identifying central figures of scientific publications. 
In this paper, we address two major limitations of their seminal contribution. First, the dataset of Yang et al.\ consists of PubMed data only, limiting the applicability of the devised supervised central figure identification model to the biomedical domain. The use of figures in scientific literature, however, is a common practice in a much broader set of research fields and areas~\cite{Lee2018}. Secondly, while supervised learning is known to generally provide the best results, it critically depends on (sufficiently) large amounts of labeled data to be used for training the models: expensive and time-consuming data annotation processes impede the scalability of central figure identification across the plethora of research domains in which figures encode valuable information. Whereas in some tasks, labeled data can be acquired more economically with crowd-sourcing, this is not the case for the task at hand: identification of the central figure for a publication requires annotators to be knowledgeable in the publication domain. 
In other words, collecting datasets large enough to support supervised learning for central figure identification for a wide range of many domains is impractical (if not infeasible) due to the high annotation costs stemming from having to recruit expert annotators.

To alleviate these issues, we propose (1) a novel benchmark for central figure identification covering several subareas of computer science, and (2) a self-supervised learning approach for which we do not need any labeled training data. For our proposed benchmark for central figure identification, we ask two (semi-expert) annotators to rank the top three figures in a scientific paper that would be the best candidates for a graphical abstract. 
The papers are collected from four computer science subdomains: natural language processing (NLP), computer vision (CV), artificial intelligence (AI), and machine learning (ML). Accordingly, our newly collected dataset allows for a comparison of the performance of central figure identification models across diverse (sub)domains.
Secondly, to eliminate the reliance on labeled training data, we introduce a self-supervised learning approach for automatic identification of central figures in scientific publications. The core idea of our approach is outlined as follows.
In most scientific publications, a figure is mentioned in an article's body by using a direct reference (e.g., \emph{``In Figure 3, we illustrate $\cdots$''}). This typically means that the paragraph of the direct link sentence roughly describes in text what the figure depicts visually, i.e., that the paragraph's content is clearly associated with the content of the figure. We exploit these direct links between an article's body and use these paragraph-figure pairs as training instances for a supervised central figure identification model.
We then train several Transformer-based \cite{NIPS2017_7181} models, which take pairs of body text and figure captions as input, and we train the models to judge whether the paragraph text (from the body of the article) matches its paired figure. At inference (i.e., test) time, we rank the article's figures by (1) predicting the scores for each article figure by pairing them all with the abstract and feeding them to the model and (2) ranking the figures based on the scores output by the model reflecting their degree of match with the article's abstract. 
In contrast to sentence matching approaches \cite{bowman-etal-2015-large,Wang2017,liu-etal-2019-original,Duan2018}, which perform sentence-pair classification, we tackle a ranking problem, scoring and ordering all figures of an article given its abstract. Although self-supervised, our approach outperforms the existing fully supervised learning approach for central figure identification \cite{Yang2019Identifying} in terms of top-1 accuracy. Finally, we provide an extensive analysis of performance differences across different domains.

\section{Related Work}
While the majority of related work in scientific paper summarization has focused on automatically creating textual summaries \cite{cohan-etal-2018-discourse,cohan-goharian-2015-scientific,Qiaozhu2008,Qazvinian2008,lauscher2017university,yasunaga2019scisummnet}, only a few have investigated the creation of visual summaries, i.e., selection of images that best reflect the publication content.
Kuzi and Zhai \cite{Kuzi2019} proposed Keyword-based figure retrieval: they tackle the related problem of ranking figures from multiple papers (in ACL anthology reference corpus \cite{bird-etal-2008-acl}). The task that we tackle in this work differs in that we focus on selecting the best figure for a single publication, considering only the figures from that publication as candidates. Similarly, in \cite{Liu2014,Yu2010} the authors rank  figures from a single paper based on their importance.

In this paper, we consider the problem of automatically identifying a central figure for a paper, which would then be a candidate for the paper's visual summary, referred to as Graphical Abstract (GA) \cite{Yang2019Identifying}. Several works have focused on analyzing GAs, e.g., their use \cite{YOON20171371} and design pattern \cite{Hullman2018}.

The automatic selection of a central figure for scientific papers was first proposed by Yang et al. \cite{Yang2019Identifying}.
In their work, they built a dataset for the central figure identification from PubMed (biomedical and life science) papers.
To extend the study of central figure identification, we propose a novel dataset consisting of computer science papers from several subdomains. Yang et al. proposed a supervised learning approach for central figure identification, a methodology that can hardly scale across a variety of scientific disciplines, due to the need for expert annotation of central figures. Limited sizes of existing datasets for various tasks in scientific publication mining \cite{Yang2019Identifying,lauscher-etal-2018-investigating,yasunaga2019scisummnet,hua-etal-2019-argument}, additionally suggest that obtaining any kind of gold expert annotations on scientific text is expensive and time-consuming. 
To remedy for this bottleneck of annotation cost, we propose a self-supervised approach in which we make use of direct inline figure references in the article body to heuristically pair article paragraphs with figure captions and use those pairs as distant supervision.

The similarity between an abstract and a figure caption is the most important feature for the supervised model of \cite{Yang2019Identifying}. Accordingly, we treat the task of identifying a central figure as an abstract-to-caption matching problem \cite{Wang2017}. Approaches for sentence matching can be divided into two types: a sentence encoding-based approach and an attention-based approach. In the sentence encoding approach, sentences are encoded separately \cite{bowman-etal-2015-large}, which, in contrast to the attention-based approach \cite{Duan2018,liu-etal-2019-original,Wang2017}, does not capture semantic interactions between them. We employ an attention-based approach and build the model on top of pretrained Transformer networks \cite{beltagy-etal-2019-scibert,devlin-etal-2019-bert}.
In contrast to current research on sentence matching as a classification task, we treat central figure identification as a ranking problem where all figures in a paper are scored according to their suitability to be used as a central figure.

\section{Annotation Study}
\label{sec:data}

\noindent
\textbf{Data Collection.} According to \cite{Lee2018}, the use of figures in scientific literature differs according to the field and the topic of research. To investigate fine-grained differences for automatic central figure identification across research domains, we collect papers published between $2017$ and $2019$ for four different research fields in computer science, namely natural language processing (NLP), computer vision (CV), artificial intelligence (AI) and machine learning (ML). In order to make the dataset sufficiently challenging, we keep only the publications with more than five figures.
Table \ref{tab:stat} provides the dataset statistics (number of publications and average number of figures per publication for each subdomain).

\setlength{\tabcolsep}{10pt}
\begin{table}[t]
\centering
\caption{Number of annotated papers per computer science sub-domain.}
\label{tab:stat}
\begin{tabularx}{\linewidth}{lccccc}
\toprule
Domain                 & NLP        & CV     & AI            & ML              & Total \\ \hline
\multirow{2}{*}{Conferences}& ACL   & \multirow{2}{*}{CVPR} & \ AAAI & \ \multirow{2}{*}{ICML}  &       \\
                       &EMNLP       &         &IJCAI          &          & \\
No.\,papers           &148         &158     &147            &144              &597     \\ 
Two annotators  &126         &127     &120            &123              &496     \\
Single annotator      &22          &31      &27             &21               &101     \\ \hline
\textit{Figures\,/\,paper} &            &        &               &                 &        \\
Average                & 6.2$\pm$1.8 & 7.0$\pm$1.8 & 6.10$\pm$1.5 & 6.5$\pm$1.9 & 6.5$\pm$1.8\\
Minimum                &5           &5       &5              &5                &5       \\
Maximum                &13          &13      &14             &13               &14      \\
\bottomrule
\end{tabularx}
\end{table}

\vspace{1em}
\noindent
\textbf{Annotation Process.} Our annotation task is defined as follows: \emph{given a paper abstract and the figures extracted from the paper, identify and rank the top 3 figures according to the degree to which they match the abstract and can therefore serve as a visual summary}. In our annotation guidelines
we adopt the definition of a graphical abstract (GA) as given in the Elsevier author guidelines (cf.\ footnote 1). Annotations were carried out by two coders with a university degree in computer science, who were instructed to study the examples provided on the publisher page and discuss them in a group to make sure they understood the notion of a graphical abstract.

To facilitate the annotation process, we develop a web-based annotation tool with a graphical user interface displaying a paper abstract and all figures extracted from the same paper, which are randomly shuffled to avoid the bias induced by the order. We first asked our annotators to read the abstract in order to obtain an overview of the paper and then to study each figure carefully. Next, the annotators were asked to choose and rank the top 3 GA candidates.
All instances are either doubly or singly annotated, and the inter-annotator agreement across the doubly annotated data amounts to $.43$ Krippendorff's $\alpha$ (ordinal), which reflects the difficulty and the subjective nature of the task.

\section{Methodology}

\noindent
\textbf{Problem Definition.} Central figure identification can be defined in two different ways \cite{Yang2019Identifying}, namely figure-level and the paper-level: in the figure-level setting, each individual figure is classified as being a central figure or not, while in the paper-level setting, a central figure is determined from all figures in a paper.
In this work, we are primarily interested in retrieving GAs as a form of summarization: hence, we opt for a document-level approach and cast it a ranking problem in which all figures from a paper are to be scored based on their suitability to provide a central figure for the publication.

Building on the result from \cite{Yang2019Identifying} that the similarity between an abstract and a figure caption is the most important factor for central figure identification, we use a pair of abstract and figure caption as input.
Given the sets of figures extracted from a paper $X=\{x_i:i\}$ and an abstract $y$, we learn a scoring function $f(x,y)$ that predicts the appropriateness of the figure to act as the central figure for the abstract (and accordingly, the paper). All figures are then ranked according to the model's prediction $S=\{s_i:s_i=f(x_i,y)\}$.

\begin{figure*}[t]
    \centering
    \includegraphics[width=0.85\linewidth]{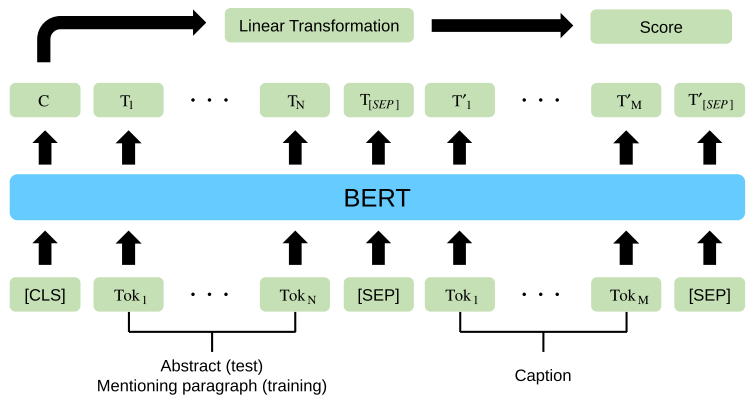}
    \caption{Our model for abstract-caption pair scoring. Paragraphs explicitly mentioning figures are paired with the figure captions during training.}
    \label{fig:model}
\end{figure*}

\vspace{1em}
\noindent
\textbf{Model.}
Our model consists of two components, a Transformer \cite{NIPS2017_7181} as a language encoder and a score predictor (Figure \ref{fig:model}).
We build upon finding from recent work in NLP that has shown the benefits of an attention-based approach for sentence matching \cite{liu-etal-2019-original,Duan2018,Wang2017}, and accordingly opt for a pre-trained BERT \cite{devlin-etal-2019-bert} model as the text encoder: specifically, we use in our experiments a SciBERT model \cite{beltagy-etal-2019-scibert}, which is pretrained on scientific publications from Semantic Scholar \cite{ammar-etal-2018-construction}. We provide a text pair consisting from an abstract and a figure caption\footnote{We feed the abstract sentences as input only at inference time. In training, input instances couple the parapgraphs from the article's body explicitly mentioning the figure with the figure caption.} as input to BERT, augmented with Transformer's special tokens: ``\texttt{[CLS] abstract [SEP] caption [SEP]}''. The transformed hidden vector of the sequence start token \texttt{[CLS]} token, $\mathbf{x}_{\mathit{CLS}}$, is then forwarded into a linear transformation layer that produces the final relevance score: $s = \mathbf{x}_{\mathit{CLS}}\mathbf{W} + b$, with the vector $\mathbf{W}\in \mathbb{R}^H$ and scalar $b\in \mathbb{R}$ as regressor's parameters ($H = 768$ is BERT's hidden state size). 
For BERT, the length of the input sequence is restricted to be up to a maximum of $512$ tokens. To overcome this limitation and allow for abstracts of arbitrary sizes, we divide an abstract into sentences and aggregate scores across sentences. Given a function $g$ to score pairs of sentences (from the abstract) and figure captions ($x$), and a set of sentences in an abstract $Y=\{y_i:i\}$, the scoring function is defined as $f(x,y) = \sum_{i} g(x,y_i)$.

\begin{figure}[t]
    \centering
    \includegraphics[width=0.85\linewidth]{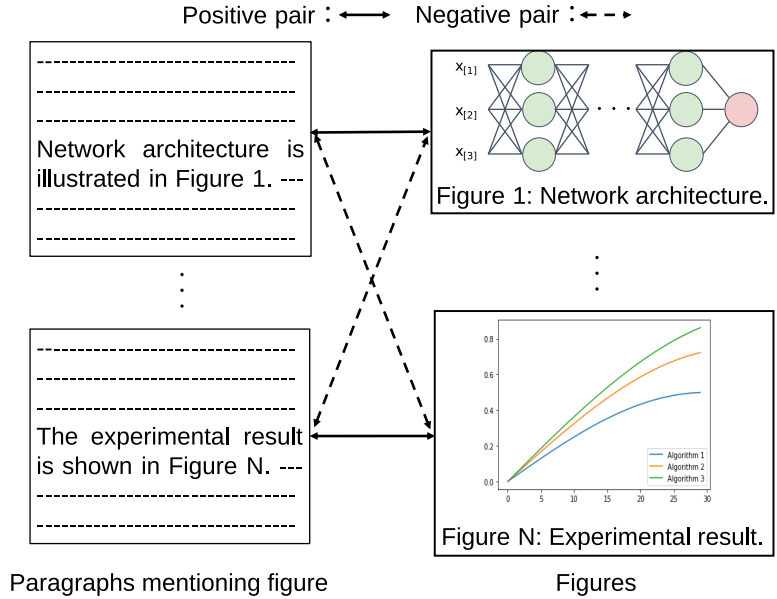}
    \caption{Creation of paragraph-figure pairs used as training instances for our models.}
    \label{fig:data}
\end{figure}

\vspace{1em}
\noindent
\textbf{Training Instance Creation.} The annotation of scientific publications requires expert knowledge of the field of research. To avoid manual annotation of the training data, we introduce a self-supervised approach by leveraging explicit inline references to figures (e.g., \textit{``Figure 2 depicts the results of the ablation experiments\dots''}).
In a scientific publication, an inline reference to a figure indicates that the paragraph and the figure are related to each other.
We denote the $i$-th paragraph that mentions the figure $x_j$ and the set of paragraphs referring to figures in a paper as $d_{i}^{j}$ and $D=\{d_{i}^{j}:i\}$, respectively.
Instead of directly identifying a central figure during training, we learn the matching of the figure $x$ and the paragraph $d$.
At training time, we make positive and negative pairs of paragraphs and figures as $(x_i,d_{j}^{k})$, as shown in Figure \ref{fig:data}.
We treat the pair $(x_i,d_{j}^{k})$ as positive if $i=k$, while $i\neq k$ for a negative one.

\vspace{1em}
\noindent
\textbf{Optimization.} We train the model to rank the positive pairs higher than negative ones.
Due to BERT's input token sequence length restriction, we randomly sample one sentence from the paragraph.
The pair of a sampled sentence and a caption is fed into the model as "[CLS] sentence [SEP] caption [SEP]".
For the training objective, we formulate the following loss similar to the Triplet loss \cite{Elad2015} as ${\cal L} = \max ( s_p - s_n + \alpha, 0)$,
where $s_p$ and $s_n$ denote the predictions of the model for the positive and negative pairs, respectively.
In the experiments, we set $\alpha = 1.0$.
For a single training instance, we sample the positive and negative pairs including the figure $x_i$ as $(x_i,d_{j}^{i})$ and $(x_i,d_{k}^{i'})$ $(i\neq i')$, respectively.
The training objective makes the score for a positive pair lower than that for a negative one: therefore, the figure with the lower score is ranked higher.

\section{Experiments}
\subsection{Implementation Details}
We conduct our experiments using
BERT's implementation from the Hugging Face library \cite{Wolf2019HuggingFacesTS}.
In all fine-tuning procedures, we use the Adam optimizer \cite{kingma2014adam} with the learning rate $1e-6$, train in batches of size $32$, and apply a dropout at the rate of $0.2$ and a gradient clipping threshold of $5$. We train the model for 1 epoch. To extract text from collected PDF versions of papers, we rely on the Science Parse library\footnote{https://github.com/allenai/science-parse}.
Explicit inline references of figure are identified via the keywords "Figure" or "Fig.". To extract figure captions, we employ the image-based approach from \cite{Siegel2018}.

\subsection{Experimental Setting}

\noindent
\textbf{PubMed.} In \cite{Yang2019Identifying}, $7,295$ biomedical and life science papers from PubMed are annotated for central figure identification. We managed to obtain the PDFs from PubMed for $7,113$ of those papers. We used the training, validation, and test splits provided by Yang et al. Using our figure mention heuristic, we created $40k$ paragraph-figure pairs from the training portion of the dataset.
Following \cite{Yang2019Identifying}, we use the top-1 and top-3 accuracy as evaluation metrics.  

\vspace{1em}
\noindent
\textbf{Computer Science (CS).} We additionally evaluate our model on our new CS dataset (Section \ref{sec:data}). Unlike the PubMed dataset, in which only a single figure is annotated as central, our annotators ranked three figures for each CS paper. Consequently, we use Mean Average Precision (MAP) and Mean Reciprocal Rank (MRR) as our evaluation metrics on the CS dataset. 
For the training, we collect papers from the same subdomains as the annotated test data, from between $2015$ and $2018$. Here we also obtain around $40k$ paragraph-figure instances for model training.

\subsection{Experiments}

\noindent
\textbf{Performance on the PubMed dataset.} We first evaluate our self-supervised approach using the PubMed dataset (Table \ref{tab:pubmed_result}).
We follow \cite{Yang2019Identifying} and make use of two baselines: random and `select first image'. For comparison, we also provide the results of two methods from \cite{Yang2019Identifying}, a text-only model that uses cosine similarity of TF-IDF between the abstract and the figure caption as the input feature, and a full model that takes the figure type label (e.g., diagram, plot) and layout (e.g., section index, figure order) as inputs, as well as text features.
We compare these against the the performance of our models based on three different pretrained Transformers, namely vanilla BERT \cite{devlin-etal-2019-bert}, RoBERTa \cite{liu2019roberta} and SciBERT \cite{beltagy-etal-2019-scibert}.

\setlength{\tabcolsep}{16pt}
\begin{table}[t]
\centering
\caption{Performance evaluation on the PubMed dataset \cite{Yang2019Identifying}.}
\label{tab:pubmed_result}
\begin{tabularx}{\linewidth}{llcc}
\toprule

Method & Model & Accuracy@1 & Accuracy@3 \\ \hline
\multirow{2}{*}{Baseline}&Random  &  0.280     &   0.701    \\
&Pick first    &  0.301     &   0.733    \\\hline
\multirow{2}{*}{Yang et al. \cite{Yang2019Identifying}}&Text-only    &  0.333     &  \textbf{0.810}     \\
&Full         &  0.344     &  0.793     \\ \hline
\multirow{3}{*}{Ours}&Vanilla BERT &  0.331     &  0.770     \\
&RoBERTa   &  0.347     & 0.741       \\
&SciBERT    &  \textbf{0.383}     &  0.787     \\
\bottomrule

\end{tabularx}
\end{table}

Regardless of the text encoder, our approach outperforms the baselines in terms of both top-1 and top-3 accuracy.
This result indicates that our method for generating  training data creation is effective for central figure identification.
Among the text encoders, SciBERT performs the best for both metrics, arguably because it has been trained on a corpus of scientific papers, thus minimizing problems related to domain transfer.
Despite not requiring manual annotation for training, our approach with SciBERT also outperforms the supervised approach of Yang et al. \cite{Yang2019Identifying} in terms of top-1 accuracy.

\vspace{1em}
\noindent
\textbf{Performance on the CS dataset.} We also evaluate the performance of the model on our CS dataset (Table \ref{tab:cs_result}). We follow the same setting as for the PubMed data and use a random and `choose first image' methods as baselines.
Here, we compare SciBERT with vanilla BERT and RoBERTa, so as to additionally verify the effectiveness of SciBERT in the CS domain -- since over $80\%$ of the papers in the corpus for SciBERT pre-training are from the biomedical domain and the ratio of CS papers account for only $18\%$ of SciBERT's pretraining corpus \cite{beltagy-etal-2019-scibert}.

\setlength{\tabcolsep}{24pt}
\begin{table}[t]
\centering
\caption{Performance evaluation on CS papers.}
\label{tab:cs_result}
\begin{tabularx}{\linewidth}{llcc}
\toprule
Method & Model       & MAP & MRR \\ \hline
\multirow{2}{*}{Baseline}&Random &  0.616     &   0.693    \\
&Pick first &  0.754     &   0.827    \\\hline
\multirow{3}{*}{Ours}&Vanilla BERT   &  0.694     &  0.773     \\
&RoBERTa    &  0.702     &  0.793     \\
&SciBERT    &  0.731     &  0.822     \\ 
\bottomrule
\end{tabularx}
\end{table}

Our approach outperforms the random baseline both in terms of MAP and MRR.
Among base models, SciBERT outperforms vanilla BERT and RoBERTa as in PubMed papers.
Though most of the corpus for SciBERT pre-training is from the biomedical domain, a certain number of CS papers seen in pretraining still contributes to the downstream performance on central figure identification.

However, as opposed to the case of PubMed papers, the `pick first' baseline here is much stronger and hard-to-beat, even for our Transformer-based approach.
This result indicates that CS papers tend to use Graphical Abstract (GA) in the beginning, and empirically highlights that our new dataset is more challenging than the PubMed-based dataset.

\vspace{1em}
\noindent
\textbf{Cross-domain Experiments.} Image usage in scientific publications is known to be different across scientific fields \cite{Lee2018}. Accordingly, we set next to empirically evaluate the robustness of our approach in a domain transfer setup. Due to the different granularity of our PubMed and CS datasets, the latter including papers from four different research areas of computer science (AI, NLP, ML, and CV), we are able to perform two sets of domain transfer experiments, namely biomedical vs.\ computer science as well as across different CS subdomains.

\setlength{\tabcolsep}{6pt}

\begin{table}[t]
\centering
\caption{Performance comparison of training on different research fields.}
\label{tab:field}
\begin{minipage}[t]{.45\textwidth}
\centering
(a) Test on PubMed papers.
\begin{tabular}{c|cc}
\toprule
Training data & ACC@1 & ACC@3 \\ \hline
-- (Random b.)     & 0.280      & 0.701     \\ \hline
PubMed   & 0.383      & 0.787      \\
CS    & 0.368      &  0.777     \\ 
\bottomrule
\end{tabular}
\\
\end{minipage}
\begin{minipage}[t]{.45\textwidth}
\centering
(b) Test on CS papers.
\\
\begin{tabular}{c|cc}
\toprule
Training data       & MAP & MRR \\ \hline
-- (Random b.)     & 0.616      & 0.693     \\ \hline
PubMed   & 0.728     & 0.822      \\
CS    & 0.731      &  0.822     \\ 
\bottomrule
\end{tabular}
\end{minipage}
\end{table}

We first compare model performance by training and testing on datasets from different domains -- i.e., biomedical papers from PubMed vs.\ computer science publications from our CS dataset -- using SciBERT as a base model (Table \ref{tab:field}).

In the test with PubMed papers, training on the same domain performs better both in terms of top-1 and top-3 accuracy.
Despite the slightly lower performance, training on the CS domain also outperforms the random baseline.
On the other hand, training on different domains, somewhat surprisingly, does not degrade the performance.
This would imply that papers from different domains exhibit similar text-figure (caption) matching properties.

\begin{table*}[t]
\centering
\caption{Performance comparison of the model trained on papers from different research topics in CS.}
\label{tab:topic}
(a) MAP
\\
\begin{tabular}{c|cccc|cc}
\toprule
\multirow{2}{*}{Test data} & \multicolumn{4}{c|}{Training data} & \multicolumn{2}{c}{Baseline} \\ \cline{2-7}
                           & NLP     & CV     & AI     & ML     &  Random &Pick first                                         \\ \hline
NLP                        & 0.727        & 0.728       & 0.728       & 0.730       &  0.631  &0.751                                \\
CV                         & 0.716        & 0.721       & 0.716       & 0.719       & 0.585   &0.758                                \\
AI                         & 0.727        &  0.729      & 0.728       & 0.730       & 0.637    &0.776                               \\
ML                         & 0.676        &  0.682      &  0.679      & 0.681       & 0.617    &0.732                               \\ 
\bottomrule
\end{tabular}
\\
\vspace{5pt}
(b) MRR
\\
\begin{tabular}{c|cccc|cc}
\toprule
\multirow{2}{*}{Test data} & \multicolumn{4}{c|}{Training data} & \multicolumn{2}{c}{Baseline} \\ \cline{2-7}
                           & NLP     & CV     & AI     & ML     & Random &Pick first                                  \\ \hline
NLP                        & 0.791        & 0.795       & 0.795       & 0.798       &  0.705  &0.816                                \\
CV                         & 0.826        & 0.833       & 0.826       & 0.831       & 0.664   &0.831                                  \\
AI                         & 0.828        & 0.834       & 0.830       & 0.828       & 0.711   &0.847                                  \\
ML                         & 0.759        & 0.769       &  0.763      & 0.769       & 0.686   &0.814                                  \\ 
\bottomrule
\end{tabular}
\end{table*}

Next, we examine the results of domain transfer  for different CS subdomains dataset belonging to different areas of computer science, due to the fact that image usage and volume may potentially vary  among fields like, e.g., natural language processing and computer vision, with papers from the latter containing typically more images.
We train models on four different areas (NLP, CV, AI, and ML) and test them on all others. Domain comparison within several research topics in CS is summarized in Table \ref{tab:topic}.

Overall, the results are rather consistent across areas and indicate that, within computer science, the research topics of papers do not affect the model performance.
Among the four topics, the performance is the lowest on machine learning (ML) papers.
One possible explanation is that many ML papers tend to describe the research from a more theoretical perspective, which does not require the use of a Graphical Abstract. The `pick first' scores higher for CV papers than for NLP and ML papers, whereas the random baseline naturally performs worse on CV papers, which contain more figures.

\begin{figure}[t]
    \centering
    \includegraphics[width=\linewidth]{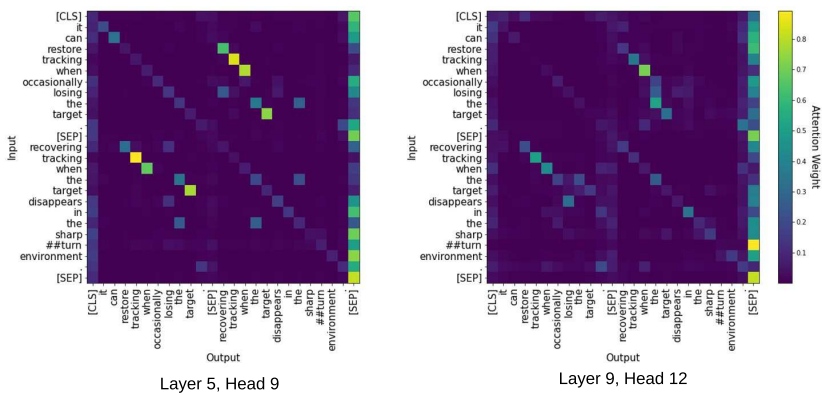}
    \caption{Examples of attentions from SciBERT which attends to semantically similar tokens (training data is from the CS domain).}
    \label{fig:attention}
\end{figure}

\vspace{1em}
\noindent
\textbf{Model Analysis.} To understand the model behavior, we analyze the attention in SciBERT.
We visualize the attention in the Transformer model.
We find that most attention maps are consistent with typical classes reported in \cite{kovaleva-etal-2019-revealing}, such as vertical or diagonal attention patterns. In some attention heads, the model attends to the lexical overlap between abstract and caption.
The examples of attention matrices produced by heads attending over the same or semantically similar tokens, are shown in Figure \ref{fig:attention}.
In this example, instances of tokens like 'tracking' and 'when' appearing in both in abstract and caption have mutually high attention weights. Additionally, pairs of tokens with similar meaning like 'restore' and 'recovering' also receive high mutual attention weights.

We also compare attention patterns among the model trained with different topics (NLP, CV, AI, and ML).
Following \cite{kovaleva-etal-2019-revealing}, we calculate the cosine similarity of attention maps.
We show the mean cosine similarity of flattened attention map for randomly selected $100$ samples in Table \ref{tab:attention}. 
Cosine similarity is high for all combinations; this means that the attention patterns across the different CS domains are virtually identical, confirming empirically our previous assumption that there are no relevant differences between CS domains when it comes to text-figure matching (see Table \ref{tab:topic}).

\begin{table}[t]
\centering
\caption{Cosine similarities of attention weight maps obtained from models trained for different CS domains. Attention maps from all layers are flattened into a single vector.}
\begin{tabular}{c|ccc}
\toprule
   & NLP & CV & AI  \\ \hline
CV & 0.9997    &  0.9998  & 0.9998       \\
AI & 0.9998    &  0.9998  & -       \\
ML & 0.9997     & -   & -       \\
\bottomrule
\end{tabular}
\label{tab:attention}
\end{table}

Kovaleva et al. reported that after fine-tuning attention maps change the most in the last two transformer layers \cite{kovaleva-etal-2019-revealing}. We therefore analyze the change in attention patterns after our task-specific fine-tuning.
We compare the standard fine-tuning, in which we update all SciBERT's parameters (and which we used in all our previous experiments), and the feature-based training, in which we freeze SciBERT's parameters and train only the regressor's parameters. The comparison of attention patterns between fine-tuned and frozen SciBERT is summarized in Table \ref{tab:finetuning}.
On the one hand, if we freeze SciBERT's parameters, we observe a major drop in performance (6 MAP points). On the other hand, high cosine similarity of attention maps between the fine-tuned and frozen SciBERT that the two transformers still exhibit similar attention patterns.
This suggests that only slight changes in the parameters of Transformer's attention heads have the potential to substantially change the predictions of the regressor.

\begin{table}[t]
    \centering
    \caption{Comparison between fine-tuned and frozen SciBERT trained on CS papers.}
    \centering
    (a) Model performance. \\
    \begin{tabular}{c|cc}
    \toprule
    SciBERT   & MAP & MRR \\ \hline
    fine-tune & 0.733    & 0.827    \\
    freeze    & 0.677    & 0.752    \\ 
    \bottomrule
    \end{tabular}
    \vspace{5pt}

    \raggedright
    (b) Cosine similarity of attention map for randomly sampled 100 sentence-caption pairs in each layer.\\
    \centering
    \begin{tabular}{c|cccccc}
    \toprule
    Layer   & 1 & 2 & 3 & 4 & 5 & 6 \\ 
    Similarity    & 0.9999    & 0.9993    &0.9983    &0.9982    &0.9983    &0.9980        \\ \hline
    Layer& 7 & 8 & 9 & 10 & 11 & 12 \\ 
    Similarity&0.9977    &0.9971    &0.9964    &0.9951    &0.9948    &0.9945 \\ 
    \bottomrule
    \end{tabular}
    \label{tab:finetuning}
\end{table}

\section{Conclusion}
While research efforts have been mostly spent on increasing information access to scientific literature by creating textual summaries, it is known that a large amount of information is often conveyed visually in the form of figures. In this work, we have addressed the problem of central figure identification from scientific publications, the task of identifying a candidate for a visual summary. Starting from  previous work, which has introduced a dataset enabling supervised learning for central figure identification, we identified and addressed two main issues: (1) the only existing data set is limited to the biomedical domain, and (2) large-scale annotations for new domains are impractical and costly. To alleviate these issues, we first presented a new benchmark collection of scientific publications annotated for central figures in the computer science domain covering four different subfields. Secondly, we proposed a self-supervised approach to central figure identification. Our method exploits the link between portions of text explicitly referencing figures and figure captions, thereby bypassing the need for large manually annotated training data. We have experimentally demonstrated the effectiveness of our approach, outperforming the supervised approach in terms of rank-1 accuracy. Finally, a follow-up analysis of cross-domain performance differences and models' attention scores revealed only slight differences across the individual CS subdomains, but interestingly, our findings also indicate that the positioning of the central figure differs between the CS and the biomedical domain. We hope that our results fuel further research on cost-effective visual summary creation for increased information access in light of the exponentially growing body of scientific literature.

\section*{Acknowledgment}
This work was supported by the Program for Leading Graduate Schools, "Graduate Program for Embodiment Informatics" of the Ministry of Education, Culture, Sports, Science and Technology (MEXT) of Japan and JST ACCEL (JPMJAC1602).
Computational resource of AI Bridging Cloud Infrastructure (ABCI) provided by National Institute of Advanced Industrial Science and Technology (AIST) was used.
%
%
%
\bibliographystyle{splncs04}
\bibliography{00-main}

\end{document}